\documentclass[superscriptaddress,twocolumn,showpacs,amsmath,amssymb]{revtex4}
\usepackage{graphicx}

\begin{document}

\title{Odd-even staggering 
of reaction cross sections for $^{22,23,24}$O isotopes
}

\author{K. Hagino}
\affiliation{ 
Department of Physics, Tohoku University, Sendai, 980-8578,  Japan} 
%E-mail:hagino@nucl.phygs.tohoku.ac.jp

\author{H. Sagawa}
\affiliation{
Center for Mathematics and Physics,  University of Aizu, 
Aizu-Wakamatsu, Fukushima 965-8560,  Japan}
%E-mail: sagawa@u-aizu.ac.jp


\begin{abstract}
The interaction cross sections of $^{22,23}$O nuclei at 900 MeV/nucleon 
have been measured recently by Kanungo {\it et al.}. 
We show that the odd-even staggering 
parameter of interaction cross sections deduced from these new data 
agrees well with the theoretical systematics expected for the neutron 
separation energy of $S_n$=2.74$\pm$0.120 MeV for $^{23}$O. 
We also discuss briefly the energy dependence of the staggering parameter. 
\end{abstract}


\maketitle

Interaction cross sections $\sigma_I$ as well as reaction 
cross sections $\sigma_R$ have been measured for many 
unstable nuclei far from the $\beta$-stability line. 
The nuclear size has been deduced from these measurements\cite{OST01}, 
and it has been revealed that the root-mean-square radii of unstable 
nuclei are significantly larger than the systematics known 
for stable nuclei. 
Particularly, 
a largely extended spatial structure, referred to as ``halo'', 
has been found for light neutron-rich nuclei close to 
the neutron drip-line \cite{Tani85,Mitt87,Ozawa01}. 

The experimental interaction cross sections for neutron-rich nuclei 
often show a large odd-even staggering (OES). That is, 
the cross section for an odd-mass nucleus is significantly larger than 
the cross sections for the neighboring even-mass nuclei. 
A typical example is the interaction cross sections for $^{30,31,32}$Ne, 
measured recently by Takechi {\it et al.}\cite{Takechi12}. 
In Ref. \cite{HS11}, we have argued that 
these large OES can be attributed to the pairing correlation. 
That is,  in odd-mass nuclei, the increase of radius of an orbit 
for an unpaired nucleon with 
a small angular momentum $l$ is 
largely suppressed 
by the pairing correlation in the even-mass nuclei. 
We have introduced the staggering parameter and shown that 
it increases for $l$=0 or 1 as the neutron separation energy, $S_n$, 
decreases. 
We have also shown that the large OES extracted from the 
experimental interaction cross sections for $^{30,31,32}$Ne is 
consistent with the theoretical systematics for $S_n$=0.29$\pm$1.64 MeV 
for $^{31}$Ne \cite{J07}. 

Recently, new measurements of the interaction cross sections for 
$^{22,23}$O were performed by Kanungo {\it et al.}\cite{K11}. 
The new data for the $^{23}$O nucleus 
is significantly smaller than the previous 
measurement by Ozawa {\it al.}\cite{Ozawa01}, which had shown 
an anomalously large cross section. 
It is also shown that 
the matter radii extracted from these new data are consistent 
with the prediction of the {\it ab initio} coupled-cluster 
theory \cite{K11}. It is therefore of interest to investigate how the 
experimental OES parameter for $^{22-24}$O 
extracted from the new data is compatible 
with the theoretical systematics shown in Ref. \cite{HS11}. 

The aim of this paper is to discuss the OES parameter of reaction cross 
sections for the $^{22,23,24}$O. 
Notice that, for neutron-rich nuclei, cross sections for inelastic 
scattering are expected to be negligibly small\cite{OYS92,KIO08}, 
and the interaction cross sections are almost the same 
as the reaction cross sections. 
Since
the reaction cross sections are much easier to calculate theoretically 
than the interaction cross sections, we consider in this paper 
the reaction cross sections for our 
analysis. 
In the previous study, we considered the $^{22,23,24}$O + $^{12}$C reactions 
at $E=240$ MeV/nucleon \cite{HS11}. 
Since the experiments by Kanungo {\it et al.} 
were performed at 
$E=900$ MeV/nucleon\cite{K11}, 
we will first discuss the energy dependence of the OES 
parameter. 

\begin{figure} 
\includegraphics[scale=0.4,clip]{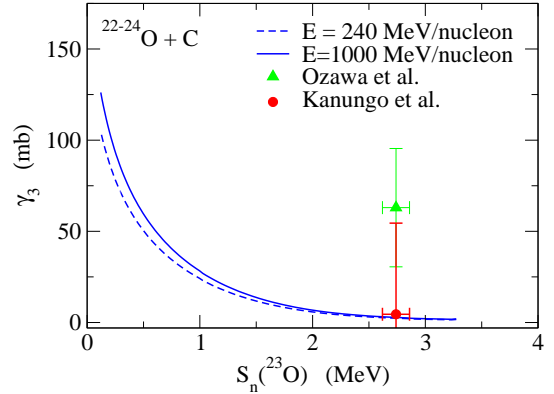}
\caption{(Color online) 
The staggering parameter $\gamma_3$ 
for $^{22,23,24}$O+$^{12}$C reactions 
as a function of 
the neutron separation energy S$_n$ for the $^{23}$O nucleus. 
The solid and the dashed lines correspond to the staggering 
parameters for the incident energies of $E$ = 1000 MeV/nucleon and 
$E$=240 MeV/nucleon, respectively. 
The filled triangle and the filled circle are the experimental 
staggering parameters extracted with the experimental data of 
Refs. \cite{Ozawa01} and \cite{K11}, respectively, plotted at the 
empirical separation energy, $S_n$=
2.74 $\pm 0.120$ MeV. 
}
\end{figure}

Figure 1 shows the OES parameter defined as \cite{HS11} 
\begin{equation}
\gamma_3=-\frac{\sigma_R(^{24}{\rm O})-2\sigma_R(^{23}{\rm O})
+\sigma_R(^{22}{\rm O})}{2},
\end{equation}
for these systems,  
where 
the reaction cross sections $\sigma_R$ are calculated with a Glauber theory. 
We use the optical limit approximation of the Glauber theory, supplemented by 
the higher order corrections \cite{AIS00}. 
The density distribution for the $^{24}$O nucleus used in the 
Glauber calculation is constructed with a 
Hartree-Fock-Bogoliubov method 
by using a Woods-Saxon mean-field potential 
together with a density-dependent zero-range pairing interaction. 
The density distributions for $^{22,23}$O, on the other hand, are 
constructed without taking into account the pairing interaction with the 
same Woods-Saxon mean-field potential, with which 
the valence neutron in $^{23}$O occupies the 2s$_{1/2}$ state. 
The OES parameter is plotted in the figure 
as a function of the one neutron separation 
energy of $^{23}$O. To this end, we vary the depth of the Woods-Saxon 
potential for the s$_{1/2}$ states. 
See Ref. \cite{HS11} for details of the calculations. 

The solid line in Fig. 1 shows the OES parameter for $E=1000$ MeV/nucleon 
while the dashed line shows that for $E=240$ MeV/nucleon. 
We use the parameters given in Table I in Ref. \cite{AIHKS08} 
for the nucleon-nucleon profile function $\Gamma_{NN}$ at each energy, 
that is related to the nucleon-nucleon scattering cross section. 
One can see that the energy dependence of the OES parameter is 
rather weak. The OES parameters at the two energies 
behave similarly to each other, although 
there exists a small deviation at small binding energies. 
This clearly indicates that the staggering parameter $\gamma_3$ provides 
a good measure for the OES of reaction cross sections, which will 
shed light on the pairing correlations in weakly bound nuclei. 

Let us now compare the theoretical curves with 
the experimental staggering parameter. 
With the experimental data of Ozawa {\it et al.} \cite{Ozawa01}, 
the staggering parameter is extracted to be 
$\gamma_3=63\pm32$ mb. 
This value is plotted in Fig. 1 by the filled triangle for a separation 
energy of $S_n$= 2.74 $\pm$ 0.120 MeV for $^{23}$O \cite{AW95}. 
One can see that the experimental staggering parameter 
$\gamma_3$ extracted from the previous 
measurement largely deviates from the theoretical systematics, for which 
the OES parameter is expected to be around 2.4 mb at $S_n\sim 2.74$ MeV. 
In marked contrast, the new data by Kanugo {\it et al.} for $^{22,23}$O
\cite{K11}, together with the previous data by Ozawa {\it et al.} 
for $^{24}$O, leads to 
$\gamma_3=4.5\pm 50.0$ mb. It is remarkable that 
this value agrees well 
 with the theoretical 
systematics, as shown by the filled circle in Fig. 1.

In summary, we have studied the odd-even staggering of reaction cross 
sections for the $^{22,23,24}$O + $^{12}$C reactions. 
We first showed that the staggering parameter depends on the incident 
energy only weakly, and thus it provides a good tool to study 
the pairing correlations in weakly bound nuclei. 
We then compared the theoretical systematics of the staggering 
parameter with the experimental values. 
The new data by 
Kanungo {\it et al.} leads to a consistent value of staggering parameter 
to the theoretical systematics, eliminating the anomaly seen in the 
previous data.  The new data are now consistent 
not only with the theoretical systematics of the 
staggering parameter, 
but also with the 
predictions of the coupled-cluster theory 
and a simple $^{22}$O+n description for $^{23}$O\cite{K11}. 
All of these 
studies
point to a conclusion that a halo structure is 
absent in the $^{23}$O nucleus. 

\bigskip

%We would like to thank M.  Takechi for fruitful discussions on 
%the experimental data. 
This work was supported by the Japanese
Ministry of Education, Culture, Sports, Science and Technology
by Grant-in-Aid for Scientific Research under
the program numbers  (C) 22540262 and  20540277.

\end{document}